\documentclass[a4paper]{jpconf}
\usepackage{graphicx}
\usepackage{amsmath}
\usepackage{amssymb}
\usepackage[utf8]{inputenc}

\def\usegraph#1#2{\includegraphics[scale=1.0,trim=0 #1 0 0]{graphs/#2.pdf}}

\def\eps{\epsilon}
\def\la{\langle}
\def\ra{\rangle}
\def\spA#1#2{\la#1#2\ra}
\def\spB#1#2{[#1#2]}

\def\al{{\alpha}}
\def\ad{{\dot{\alpha}}}

\def\MGaMC{\textsc{Madgraph5}\_a\textsc{MC@NLO}}
\def\NJet{\textsc{NJet}}
\def\BlackHat{\textsc{BlackHat}}
\def\GoSam{\textsc{GoSam}}
\def\OpenLoops{\textsc{OpenLoops}}
\def\Recola{\textsc{Recola}}

\def\Ninja{\textsc{Ninja}}

\def\Sherpa{\textsc{Sherpa}}
\def\Herwig{\textsc{Herwig7}}

\begin{document}
\title{Automating QCD amplitudes with on-shell methods}

\author{Simon Badger}

\address{Higgs Centre for Theoretical Physics, School of Physics and Astronomy, The University of Edinburgh, Edinburgh EH9 3JZ, Scotland, UK.\\}

\ead{sbadger@staffmail.ed.ac.uk}

\begin{abstract}
  We review some of the modern approaches to scattering amplitude computations in QCD
  and their application to precision LHC phenomenology. We emphasise the usefulness of momentum twistor variables
  in parameterising general amplitudes.
\end{abstract}

\section{Introduction}

New understanding of the structure of scattering amplitudes has led to
efficient algorithms for high multiplicity final states which are now common in
next-to-leading order simulations of LHC collisions. To keep up with the
experimental accuracy expected during Run II, new theoretical methods are
required to look at high multiplicity final states at higher accuracy. In this
contribution we take a brief look at some new on-shell methods attempting to
tackle this problem. As a specific example we consider applications of momentum
twistor technology for the parameterisation of multi-leg kinematics.

Over the last ten years or so an increasingly large number of theoretical
physicists have turned their attention to understanding the underlying
structure of amplitudes in gauge, gravity and string theory. Much of this can
be traced to Witten's 2003 paper on $\mathcal{N}=4$ super-symmetric Yang-Mills
theory as a string theory in twistor space~\cite{Witten:2003nn} though on-shell
unitarity methods~\cite{Bern:1994zx,Bern:1994cg} have been hard at work in collider physics applications for much longer.
Perhaps the most remarkable example of on-shell simplicity was the realisation
of Parke and Taylor~\cite{Parke:1986gb} that tree-level amplitudes for
multi-gluon scattering amplitudes with specific external helicity states can be written
in a compact form,
\begin{align}
  A_n^{(0)}(1^+,2^+,3^+,\ldots,n^+) &= 0, \\
  A_n^{(0)}(1^-,2^+,3^+,\ldots,n^+) &= 0, \\
  A_n^{(0)}(1^-,2^-,3^+,\ldots,n^+) &= \frac{i\spA12^4}{\spA12\spA23\spA34\cdots\spA{n}1}.
  \label{eq:MHVtrees}
\end{align}
This famous representation for the maximal helicity violating configuration at the amplitude level, due to later work by
Mangano, Parke and Xu~\cite{Mangano:1987xk} and Berends and
Giele~\cite{Berends:1987me}, resulted from understanding the power of both
colour ordering to exploit the $SU(N_c)$ group structure and the
spinor-helicity formalism to simplify the on-shell kinematics. The motivation
for studying these high multiplicity tree-level expressions came from collider physics
applications where multi-jet backgrounds were still impossible even at leading order
in perturbation theory.

\begin{table}[hbp]
  \centering
  \begin{tabular}{cccccccccc}
    \hline
    $n$              & 3 & 4 & 5 & 6 & 7 & 8 & 9 & 10 \\
    \hline
    Feynman diagrams & 1 & 4 & 25 & 220 & 2485 & 34300 & 559405 & 10525900 \\
    Ordered diagrams & 1 & 3 & 10 & 38 & 154 & 654 & 2871 & 12925 \\
    MHV              & 1 & 1 & 1 & 1 & 1 & 1 & 1 & 1 \\
    BCFW NMHV        & - & - & - & 3 & 6 & 10 & 15 & 21 \\
    BCFW N${}^2$MHV  & - & - & - & - & - & 20 & 50 & 105 \\
    BCFW N${}^3$MHV  & - & - & - & - & - & - & - & 175 \\
    \hline
  \end{tabular}
  \caption{The number of graphs appearing in tree-level $n$-gluon scattering using different approaches. The number of terms
  in BCFW are chosen for the most complicated alternating helicity case and were obtained using Bourjaily's {\tt bcfw} package \cite{Bourjaily:2010wh}.}
  \label{tab:diagcount}
\end{table}

Though the number of Feynman diagrams involved in a given computation are sometimes used as measure of the complexity, with
a high degree of automation they can easily be used to compute relevant processes. Obtaining compact and efficient amplitude
representations with this method is fairly hopeless however. In Table \ref{tab:diagcount}
we compare the set of Feynman graphs for the full colour amplitude $\mathcal{A}_n^{(0)}$ with the reduced colour ordered set of graphs
which contribute to the partial amplitudes $A_n^{(0)}$ defined by,
\begin{equation}
  \mathcal{A}_n^{(0)} = \sum_{\sigma\in S_{n-2}}
      \tilde{f}^{a_1 \sigma(a_2) \sigma(a_3)} \cdots
      \tilde{f}^{\sigma(a_{n-2}) \sigma(a_{n-1}) a_n}
      A_n^{(0)}(1,\sigma(2),\ldots,\sigma(3),n)
  \label{eq:ngcolordtree}
\end{equation}
Here we have used the compact colour decomposition into adjoint traces~\cite{DelDuca:1999rs} which incorporates additional colour
symmetry coming from Kleiss-Kuijf identities~\cite{Kleiss:1988ne} and runs over $(n-2)!$ permutations. $\tilde{f}^{abc} = \sqrt{2}f^{abc}$
are the usual structure constants in $SU(N_c)$. By exploiting the symmetry in colour space the amplitude can be constructed from
a minimal set of gauge invariant kinematic building blocks. The number of diagrams $N_{\mathcal{A}}(n)$ in the full colour amplitudes can be derived from a
simple algorithm using differential operators~\cite{Caravaglios:1998yr}. If we let $t$ be the number of three-gluon vertices and $g$ be the number of internal
and external gluons then,
\begin{equation}
  N_{\mathcal{A}}(n) = \left( t\,g^3\,\frac{\partial}{\partial g} +  g \,\frac{\partial}{\partial t} \right)^{n-3} t\,g^3 \Bigg|_{t=g=1},
  \label{eq:ndiagfull}
\end{equation}
where the first derivative adds a gluon to each gluon line via a three vertex and the second derivative adds a gluon
to each three vertex via a four-point vertex. The colour ordered diagrams can be quickly computed using the recursive
approach of Berends and Giele~\cite{Berends:1987me}.
\begin{equation}
  N_A(n) =
  \sum_{k=1}^{n-2} N_A(k+1) N_A(n-k) +
  \sum_{k=1}^{n-3} \sum_{l=k+1}^{n-3} N_A(k+1) N_A(l-k+1) N(n-l)
  \label{eq:ndiagord}
\end{equation}
A more modern technique for the evaluation of tree-level amplitudes purely from
on-shell building blocks is the recursive method of Britto, Cachazo, Feng and
Witten \cite{Britto:2004ap,Britto:2005fq}. By counting the number of on-shell
BCFW diagrams that contribute to a given helicity configuration, one can see
that there is considerable simplification from the ordered diagrams to a fully
on-shell amplitude. Automated packages which solve the on-shell
recursion relation to obtain compact expression for all tree amplitudes in
massless QCD have been developed~\cite{Dixon:2010ik,Bourjaily:2010wh}. The complexity of helicity
amplitudes then becomes clear with the addition of more and more negative
helicity gluons.

For applications in collider phenomenology an individual helicity amplitude is
unfortunately of little use and both sums of the helicity and colour
configurations of the ordered partial amplitudes must be performed efficiently.
Given the precision for the LHC experiments, leading order predictions in QCD
are insufficient for four or even five particle final states, so it is
necessary to perform this task at least to one-loop order.

Nevertheless a host of new algorithms have been developed in recent years which
exploit this tree-level simplicity at the loop level and can provide the necessary
information to the Monte Carlo event generators like \Sherpa~\cite{Gleisberg:2008ta}, \MGaMC~\cite{Alwall:2014hca}
and \Herwig~\cite{Bellm:2015jjp} used in experimental analyses.
A few important insights were necessary to make this possible some of which I
will expand on the next sections.

\section{Unitarity, generalised unitarity and integrand reduction}

One of the key insights which led to the automation of one-loop amplitudes is
that a purely algebraic algorithm allows the kinematic algebra to be performed
numerically and avoid a traditional bottleneck in multi-leg amplitude
computations.

The starting point for this approach is the well known decomposition of
one-loop amplitudes into a basis of known box, triangle and bubble integrals,
\begin{equation}
  A^{(1)}_n = \sum_i c_i I^{4-2\eps}_i + \text{ rational} + \mathcal{O}(\eps).
  \label{eq:1lamp}
\end{equation}
In this formula for one-loop primitive (i.e. colour ordered) amplitudes $c_i$
are rational functions while $I^{4-2\eps}_i$ are loop integrals in the
dimensional regularisation parameter $\eps$.  The generalised unitarity
algorithm uses complex momenta to systematically solve the multiple cut
conditions and isolate the coefficients $c_i$ from products of tree-level
amplitudes \cite{Britto:2004nc,Forde:2007mi,Berger:2008sj}.  The remaining
rational term can be extracted from modified BCFW recursion relations
\cite{Bern:2005hs,Bern:2005ji,Bern:2005cq,Berger:2006ci,Berger:2006vq} or
$D$-dimensional unitarity cuts \cite{Badger:2008cm}.

Another approach developed at the same time was the integrand reduction method
of Ossola, Papadopoulos and Pittau~\cite{Ossola:2006us}.  Here the loop
momentum dependence of the integrand is completely parameterised in terms of
the scalar integral basis and additional 'spurious' terms which integrate to
zero. The sum over topologies remains the same while the numerator is
decomposed into monomials of irreducible scalar products $\Delta_i(k)$:
\begin{equation}
  A^{(1)}_n = \int d^{4-2\eps} \sum_i \frac{\Delta_i(k)}{\prod_{\alpha\in i} D_\alpha(k)},
  \label{eq:1lint}
\end{equation}
where $D_\alpha(k)$ are the propagators for the $i$th topology.
The integrand reduction procedure is compatible with $D$-dimensional
generalised unitarity cuts \cite{Ellis:2007br,Giele:2008ve}. A complete
reconstruction of the integrand, including spurious terms, is important to
allow the algorithm to be implemented efficiently using fixed precision
numerics. Further details on these algorithms can be found in reference
\cite{Ellis:2011cr}.

Since the information required for any loop amplitude has now been
parameterised in terms of rational functions, the generation of the input to
these algorithms can then be performed using tree-level recursive techniques.
An important development in this area was the extension of off shell recursive
techniques to tensor integrands using the \OpenLoops~method
\cite{Cascioli:2011va}. Variations of these techniques have subsequently been
implemented into computer codes such as \GoSam~\cite{Cullen:2014yla},
\OpenLoops~\cite{Cascioli:2011va}, \textsc{MadLoop}\footnote{\textsc{MadLoop} is
part of the \MGaMC Monte Carlo event generator which offers fully automated NLO
simulations interfaced with parton showers.}~\cite{Hirschi:2011pa} and
\Recola~\cite{Actis:2012qn,Actis:2016mpe} that can handle a wide variety of complicated
processes including QCD and Electro-weak corrections
\GoSam~\cite{Cullen:2014yla} combines efficient Feynman diagram generation
together with the integrand reduction algorithm \Ninja~\cite{Peraro:2014cba}.

The \BlackHat~\cite{Berger:2008sj} and \NJet~\cite{Badger:2012pg} one-loop
amplitude providers are more specialised but use a fully on-shell approach
that is currently able to attack higher multiplicity processes than other
algorithms.  These codes have been interfaced with the \Sherpa Monte-Carlo to
produce NLO predictions for high multiplicity processes like $pp\to
5j$~\cite{Badger:2013yda}, $pp\to WW+3j$~\cite{Cordero:2015hem}, $pp\to
\gamma\gamma+3j$~\cite{Badger:2013ava} and $pp\to W+5j$~\cite{Bern:2013gka}.
The bottleneck in these extreme configurations is no longer in the virtual corrections
since most of the event generation time is spent in the real radiation phase-space
integration. Nevertheless, without the advanced recursive tree amplitude algorithms
in \textsc{Comix}~\cite{Gleisberg:2008fv} and \Sherpa's automated dipole subtraction
method~\cite{Gleisberg:2007md} such predicitions wouldn't be possible at all.

\section{$D$-dimensional generalised unitarity for multi-loop integrands}

Extending the current multi-loop methods to higher multiplicity represents
a serious challenge. The $D$-dimensional generalised unitarity cut algorithm has recently
been extended to multi-loop integrands using integrand reduction
\cite{Ossola:2006us} and elements of computational algebraic geometry~%
\cite{Mastrolia:2011pr,Badger:2012dp,Zhang:2012ce,Kleiss:2012yv,Feng:2012bm,Mastrolia:2012an,Mastrolia:2013kca,Badger:2013gxa}.
In contrast to the one-loop case, the basis of integrals obtained through this
method is not currently known analytically and is much larger than the set of
basis functions defined by standard integration-by-parts(IBP) identities.
Nevertheless, new results in non-supersymmetric theories have been obtained for
five gluon scattering amplitudes with all positive helicities
\cite{Badger:2013gxa,Badger:2015lda}.  The maximal unitarity method
\cite{Kosower:2011ty}, which incorporates IBP identites, has been applied to a
variety of two-loop examples in four dimensions
\cite{Larsen:2012sx,CaronHuot:2012ab,Johansson:2012zv,Johansson:2013sda,Johansson:2015ava}.
This approach can be seen as an extension of the generalised unitarity methods
of Britto, Cachazo and Feng~\cite{Britto:2004nc} and Forde~\cite{Forde:2007mi}.
Efficient algorithms to generate unitarity compatible IBPs are a key ingredient
in both approaches and has been the focus of on-going investigations
\cite{Gluza:2010ws,Schabinger:2011dz,Ita:2015tya,Larsen:2015ped}.

Constructing loop amplitudes from trees for non-supersymmetric theories
like QCD require the identification of the additional components of the dimensionally
regulated loop momenta. At two-loops this information can be extracted from six-dimensional
trees which are computed efficiently using the six-dimensional spinor-helicity formalism \cite{Cheung:2009dc}.
These cuts can then be dimensionally reduced to $4-2\eps$ in an analogous way to the one-loop
approaches considered previously \cite{Bern:2010qa,Davies:2011vt}.

Breaking loop amplitudes into trees also has benefits for colour decompositions and representations of the non-planer
sector since tree-level amplitude relations such as Kliess-Kuijf~\cite{Kleiss:1988ne} and Bern-Carrasco-Johansson \cite{Bern:2008qj}
can be applied at the level of the cut amplitude. This technique has been shown to be a powerful tool in $\mathcal{N}=4$ Yang-Mills and $\mathcal{N}=8$ super-gravity
computations \cite{Bern:2012uf}, but recently been exploited in the computation of the
complete five gluon all-plus amplitude at two loops in pure Yang-Mills \cite{Badger:2015lda}. Combining
$D$-dimensional generalised unitarity cuts, BCJ relations and integrand reduction a remarkably compact form
was obtained,
\begin{equation}\begin{aligned} \!\!\!\!\!\!\!\!
   {\cal A}^{(2)}(1^+\!,2^+\!,3^+\!,4^+\!,5^+) =
   \qquad \qquad \qquad \qquad \qquad \qquad
   \qquad \qquad \qquad \qquad \qquad \qquad &
   \!\!\!\!\!\!\!\!\!\! \\ \!\!\!\!\!\!\!
   ig^7 \sum_{\sigma \in S_5} \sigma \circ
   I\Bigg[\,
      C\bigg(\usegraph{9}{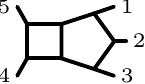}\!\!\!\:\bigg) 
      \Bigg(
      \frac{1}{2} \Delta\bigg(\usegraph{9}{delta431i}\!\!\!\:\bigg)
 \! + \Delta\bigg(\usegraph{9}{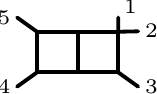}\bigg)
 \! + \frac{1}{2} \Delta\bigg(\usegraph{13}{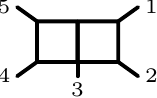}\bigg) &
      \!\!\!\!\!\!\!\!\!\! \\
 \!+\,\frac{1}{2} \Delta\bigg(\usegraph{13}{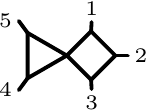}\!\!\!\:\bigg)
 \! + \Delta\bigg(\usegraph{10}{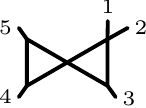}\bigg)
 \! + \frac{1}{2} \Delta\bigg(\usegraph{13}{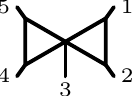}\bigg) &
 \!   \Bigg) \!\!\!\!\!\!\!\!\!\! \\ \!\!\!\!\!
    + C\bigg(\usegraph{9}{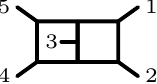}\bigg)
      \Bigg(
      \frac{1}{4} \Delta\bigg(\usegraph{9}{delta332NPi}\bigg)
 \! + \frac{1}{2} \Delta\bigg(\usegraph{9}{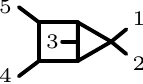}\!\bigg)
 \! + \frac{1}{2} \Delta\bigg(\usegraph{9}{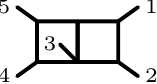}\bigg) & \!\!\!\!\!\!\!\!\!\! \\
 \!-\,\Delta\bigg(\!\!\!\;\usegraph{16}{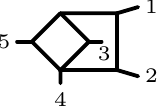}\bigg)
 \! + \frac{1}{4} \Delta\bigg(\usegraph{9}{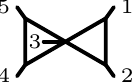}\bigg) &
 \!   \Bigg) \!\!\!\!\!\!\!\!\!\! \\
    + C\bigg(\!\!\!\:\usegraph{9}{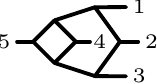}\!\!\!\:\bigg)
      \Bigg(
      \frac{1}{4} \Delta\bigg(\!\!\!\:\usegraph{9}{delta422NPi}\!\!\!\:\bigg)
 \! + \frac{1}{2} \Delta\bigg(\!\!\!\:\usegraph{8.8}{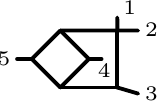}\bigg)
 \!   \Bigg)
   \Bigg] . \!\!\!\!\!\!\!\!\!\!
\label{eq:A5point2loop2}
\end{aligned}\end{equation}
where the $C$ functions are the colour factors in terms of adjoint $\tilde{f}$ structure constants
and $\Delta$ are the irreducible numerators. Further details can be found in reference \cite{Badger:2015lda}.

\section{Momentum Twistors}

In this section we collect some useful formulae for parameterising external
kinematics using Hodges' momentum twistor formalism~\cite{Hodges:2009hk}.
Momentum twistors are extremely useful for studying the geometric properties of
amplitudes but they also have a practical property that can be used in general
amplitude computations: these variable linearise the momentum conservation
conditions and give a rational parameterisation of the phase-space.

The topic is somewhat more specialised than the techniques reviewed in the previous sections.
The connection is that rational parameterisation of the kinematics can allow analytic
computations to be performed in exactly the same way as the numerical algorithms
described above and hence is a useful tool when combined with unitarity cuts and integrand reduction.

The notation used in this section relies on the spinor-helicity formalism of which good introductions
can be found in references~\cite{Elvang:2013cua,Henn:2014yza,Dixon:1996wi,Mangano:1990by}. The first observation
stems from the fact that massless momenta can be decomposed in two component Weyl spinors,
\begin{equation}
  p^2 = 0 \Rightarrow (\sigma\cdot p)^{\alpha\dot\alpha} = \lambda^\alpha(p) \tilde{\lambda}^{\dot{\alpha}}(p).
\end{equation}
Hence the on-shell conditions are manifest when using the spinor-helicity formalism. Momentum conservation,
\begin{equation}
  \sum_i p_i^\mu = 0,
  \label{eq:momcons}
\end{equation}
still impose some non-trivial conditions on the amplitudes. Hodges \cite{Hodges:2009hk} introduced momentum twistors
as a natural extension of Penrose's twistor formalism. We being by defining dual momentum co-ordinates, $x_i^\mu$,
\begin{equation}
  p^\mu_i = x^\mu_i - x^\mu_{i-1}
\end{equation}
which can be inverted up to a fixed point, $x_0^\mu$,
\begin{equation}
  x^\mu_i = x^\mu_0 + \sum_{k=1}^i p^\mu_k.
\end{equation}
The momentum twistor is then constructed from these dual co-ordinates and the holomorphic
Weyl spinors:
\begin{equation}
  Z_{iA} = \left(\lambda_\al(i), \mu^\ad(i) = \lambda_\al(i)(\sigma\cdot x_i)^{\al\ad} \right),
\end{equation}
The two component object $\mu^\ad(i)$ is used instead of the $\tilde{\lambda}^\ad(i)$ spinor
to define the kinematics for the $n$-particle system $i=1,n$.
The $\tilde{\lambda}(i)_\ad$ spinor is defined through the dual twistor,
\begin{align}
  W_{i}^A = (\tilde{\mu}_\al(i), \tilde{\lambda}^\ad(i)) = \frac{\varepsilon^{ABCD} Z_{(i-1)B} Z_{iC} Z_{(i+1)D}}{\spA{i-1}{i}\spA{i}{i+1}}
  \label{eq:dualtwistor}
\end{align}
from which we can find the definition of the anti-holomorphic spinor,
\begin{equation}
  \tilde{\lambda}(i)^\ad = \frac{
    \spA{i-1}{i}\mu^\ad(i+1) + \spA{i+1}{i-1} \mu^\ad(i) + \spA{i}{i+1}\mu^\ad(i-1)
  }{
    \spA{i-1}{i}\spA{i}{i+1}
  }.
  \label{eq:aspinordef}
\end{equation}
By use of the Schouten identity it is easy to show that any $\tilde{\lambda}(i)_\ad$ defined through
the dual twistor will automatically satisfy momentum conservation. Amplitudes can then be written in terms
of holomorphic spinor products and momentum twistor 4-brackets:
\begin{equation}
  \la i j k l \ra = \varepsilon^{ABCD} Z_{iA}Z_{jB}Z_{kC}Z_{lD}.
\end{equation}

The $4\times n$ matrix $Z_{iA}$ has $3n-10$ independent entries and we can generate rational phase-space
points (for complex momenta) by filling the matrix with random integers. We can even pick rational functions
to fill the momentum twistor matrix, which amounts to picking a special frame to evaluate general kinematics.
A convenient choice is,
\begin{equation}
  Z_{i} =
    \begin{pmatrix}
      \frac{\Sigma_i}{s_{12}} \\
      1 - \delta_{1i} \\
      \frac{\la 123i \ra\spA34\spB23}{\la 1234 \ra \spA1i\spB12} \\
      \frac{-\spA13\la 124i \ra + \spA14\la 123i \ra}{\la 1234 \ra\spA1i}
    \end{pmatrix}
  \label{eq:mtparam}
\end{equation}
where
\begin{equation}
  \frac{\Sigma_i}{s_{12}} =
  \begin{cases}
    0 & i=1 \\
    \frac{\spA13\spA2i}{\spA23\spA1i} & \text{otherwise}
  \end{cases}
\end{equation}
One additional subtlety in using this technique is that phase information ensures parity invariance is discarded in
favour of rational functions. This can be easily restored as a pre-factor. In the explicit parameterisation
above this is:
\begin{equation}
  \Phi_n(1^{h_1},\cdots,n^{h_n}) = \left(\frac{\spA13}{\spB12\spA23}\right)^{-h_1} \prod_{i=2}^n \left(\frac{\spA1i^2\spB12\spA23}{\spA13}\right)^{-h_i}
  \label{eq:MTphase}
\end{equation}
where $h_i$ are the helicities of the external gluons. As an example the MHV amplitude in this language would
read,
\begin{equation}
  \frac{A_n^{(0)}(1^-,2^-,3^+,\ldots,n^+)}{\Phi_n(1^-,2^-,3^+,\ldots,n^+)} = \frac{i\,s_{12}^{n-2}}{(\Sigma_4-1)\prod_{i=4}^{n-1}(\Sigma_i-\Sigma_{i+1})}
  \label{eq:MHVMT}
\end{equation}
The power of this representation is that the above expression can be obtained
by simply plugging in the parameterisation \eqref{eq:mtparam} into the
thousands of ordered Feynman diagrams - assuming your computer has enough
memory to perform the multi-variate factorisation. The MHV is a special case
of course since the being a single term the factorisation is guaranteed to land
on the compact expression.

\section{Conclusions}

In these proceedings we have given a very brief overview of the on-shell
methods used in automated loop amplitude computations. The success of this
approach to the one-loop problem has motivated activity in applying the methods
at higher loops and some useful progress has been made. Some subtleties of the
integrand reduction method beyond one-loop have been identified in particular
the non-minimal basis of integrals compared to those generated by traditional
IBP reduction to master integrals.

We highlighted one of the new techniques used in modern high multiplicity loop computations.
Though momentum twistors were originally applied in the context of maximally super-symmetric Yang-Mills
theory these computations show they can be applied just as easily to general gauge theories.

\section*{Acknowledegments}

It is a great pleasure to thank the organisers of ACAT 2016 for the invitation to present this work.
The work has been supported by STFC Rutherford Fellowship ST/L004925/1. Thanks also go to Donal O'Connell
and Francesco Buciuni for useful comments.

\section*{References}

\bibliographystyle{iopart-num}
\bibliography{acat2016}

\providecommand{\newblock}{}
\begin{thebibliography}{10}
\expandafter\ifx\csname url\endcsname\relax
  \def\url#1{{\tt #1}}\fi
\expandafter\ifx\csname urlprefix\endcsname\relax\def\urlprefix{URL }\fi
\providecommand{\eprint}[2][]{\url{#2}}

\bibitem{Witten:2003nn}
Witten E 2004 {\em Commun. Math. Phys.\/} {\bf 252} 189--258 (\textit{Preprint}
  \eprint{hep-th/0312171})

\bibitem{Bern:1994zx}
Bern Z, Dixon L~J, Dunbar D~C and Kosower D~A 1994 {\em Nucl. Phys.\/} {\bf
  B425} 217--260 (\textit{Preprint} \eprint{hep-ph/9403226})

\bibitem{Bern:1994cg}
Bern Z, Dixon L~J, Dunbar D~C and Kosower D~A 1995 {\em Nucl. Phys.\/} {\bf
  B435} 59--101 (\textit{Preprint} \eprint{hep-ph/9409265})

\bibitem{Parke:1986gb}
Parke S~J and Taylor T~R 1986 {\em Phys. Rev. Lett.\/} {\bf 56} 2459

\bibitem{Mangano:1987xk}
Mangano M~L, Parke S~J and Xu Z 1988 {\em Nucl. Phys.\/} {\bf B298} 653

\bibitem{Berends:1987me}
Berends F~A and Giele W~T 1988 {\em Nucl. Phys.\/} {\bf B306} 759

\bibitem{Bourjaily:2010wh}
Bourjaily J~L 2010  (\textit{Preprint} \eprint{1011.2447})

\bibitem{DelDuca:1999rs}
Del~Duca V, Dixon L~J and Maltoni F 2000 {\em Nucl. Phys.\/} {\bf B571} 51--70
  (\textit{Preprint} \eprint{hep-ph/9910563})

\bibitem{Kleiss:1988ne}
Kleiss R and Kuijf H 1989 {\em Nucl. Phys.\/} {\bf B312} 616

\bibitem{Caravaglios:1998yr}
Caravaglios F, Mangano M~L, Moretti M and Pittau R 1999 {\em Nucl. Phys.\/}
  {\bf B539} 215--232 (\textit{Preprint} \eprint{hep-ph/9807570})

\bibitem{Britto:2004ap}
Britto R, Cachazo F and Feng B 2005 {\em Nucl. Phys.\/} {\bf B715} 499--522
  (\textit{Preprint} \eprint{hep-th/0412308})

\bibitem{Britto:2005fq}
Britto R, Cachazo F, Feng B and Witten E 2005 {\em Phys. Rev. Lett.\/} {\bf 94}
  181602 (\textit{Preprint} \eprint{hep-th/0501052})

\bibitem{Dixon:2010ik}
Dixon L~J, Henn J~M, Plefka J and Schuster T 2011 {\em JHEP\/} {\bf 01} 035
  (\textit{Preprint} \eprint{1010.3991})

\bibitem{Gleisberg:2008ta}
Gleisberg T, Hoeche S, Krauss F, Schonherr M, Schumann S, Siegert F and Winter
  J 2009 {\em JHEP\/} {\bf 02} 007 (\textit{Preprint} \eprint{0811.4622})

\bibitem{Alwall:2014hca}
Alwall J, Frederix R, Frixione S, Hirschi V, Maltoni F, Mattelaer O, Shao H~S,
  Stelzer T, Torrielli P and Zaro M 2014 {\em JHEP\/} {\bf 07} 079
  (\textit{Preprint} \eprint{1405.0301})

\bibitem{Bellm:2015jjp}
Bellm J {\em et~al.\/} 2016 {\em Eur. Phys. J.\/} {\bf C76} 196
  (\textit{Preprint} \eprint{1512.01178})

\bibitem{Britto:2004nc}
Britto R, Cachazo F and Feng B 2005 {\em Nucl. Phys.\/} {\bf B725} 275--305
  (\textit{Preprint} \eprint{hep-th/0412103})

\bibitem{Forde:2007mi}
Forde D 2007 {\em Phys. Rev.\/} {\bf D75} 125019 (\textit{Preprint}
  \eprint{0704.1835})

\bibitem{Berger:2008sj}
Berger C~F, Bern Z, Dixon L~J, Febres~Cordero F, Forde D, Ita H, Kosower D~A
  and Maitre D 2008 {\em Phys. Rev.\/} {\bf D78} 036003 (\textit{Preprint}
  \eprint{0803.4180})

\bibitem{Bern:2005hs}
Bern Z, Dixon L~J and Kosower D~A 2005 {\em Phys. Rev.\/} {\bf D71} 105013
  (\textit{Preprint} \eprint{hep-th/0501240})

\bibitem{Bern:2005ji}
Bern Z, Dixon L~J and Kosower D~A 2005 {\em Phys. Rev.\/} {\bf D72} 125003
  (\textit{Preprint} \eprint{hep-ph/0505055})

\bibitem{Bern:2005cq}
Bern Z, Dixon L~J and Kosower D~A 2006 {\em Phys. Rev.\/} {\bf D73} 065013
  (\textit{Preprint} \eprint{hep-ph/0507005})

\bibitem{Berger:2006ci}
Berger C~F, Bern Z, Dixon L~J, Forde D and Kosower D~A 2006 {\em Phys. Rev.\/}
  {\bf D74} 036009 (\textit{Preprint} \eprint{hep-ph/0604195})

\bibitem{Berger:2006vq}
Berger C~F, Bern Z, Dixon L~J, Forde D and Kosower D~A 2007 {\em Phys. Rev.\/}
  {\bf D75} 016006 (\textit{Preprint} \eprint{hep-ph/0607014})

\bibitem{Badger:2008cm}
Badger S~D 2009 {\em JHEP\/} {\bf 01} 049 (\textit{Preprint}
  \eprint{0806.4600})

\bibitem{Ossola:2006us}
Ossola G, Papadopoulos C~G and Pittau R 2007 {\em Nucl. Phys.\/} {\bf B763}
  147--169 (\textit{Preprint} \eprint{hep-ph/0609007})

\bibitem{Ellis:2007br}
Ellis R~K, Giele W~T and Kunszt Z 2008 {\em JHEP\/} {\bf 03} 003
  (\textit{Preprint} \eprint{0708.2398})

\bibitem{Giele:2008ve}
Giele W~T, Kunszt Z and Melnikov K 2008 {\em JHEP\/} {\bf 04} 049
  (\textit{Preprint} \eprint{0801.2237})

\bibitem{Ellis:2011cr}
Ellis R~K, Kunszt Z, Melnikov K and Zanderighi G 2012 {\em Phys. Rept.\/} {\bf
  518} 141--250 (\textit{Preprint} \eprint{1105.4319})

\bibitem{Cascioli:2011va}
Cascioli F, Maierhofer P and Pozzorini S 2012 {\em Phys. Rev. Lett.\/} {\bf
  108} 111601 (\textit{Preprint} \eprint{1111.5206})

\bibitem{Cullen:2014yla}
Cullen G {\em et~al.\/} 2014 {\em Eur. Phys. J.\/} {\bf C74} 3001
  (\textit{Preprint} \eprint{1404.7096})

\bibitem{Hirschi:2011pa}
Hirschi V, Frederix R, Frixione S, Garzelli M~V, Maltoni F and Pittau R 2011
  {\em JHEP\/} {\bf 05} 044 (\textit{Preprint} \eprint{1103.0621})

\bibitem{Actis:2012qn}
Actis S, Denner A, Hofer L, Scharf A and Uccirati S 2013 {\em JHEP\/} {\bf 04}
  037 (\textit{Preprint} \eprint{1211.6316})

\bibitem{Actis:2016mpe}
Actis S, Denner A, Hofer L, Lang J~N, Scharf A and Uccirati S 2016
  (\textit{Preprint} \eprint{1605.01090})

\bibitem{Peraro:2014cba}
Peraro T 2014 {\em Comput. Phys. Commun.\/} {\bf 185} 2771--2797
  (\textit{Preprint} \eprint{1403.1229})

\bibitem{Badger:2012pg}
Badger S, Biedermann B, Uwer P and Yundin V 2013 {\em Comput. Phys. Commun.\/}
  {\bf 184} 1981--1998 (\textit{Preprint} \eprint{1209.0100})

\bibitem{Badger:2013yda}
Badger S, Biedermann B, Uwer P and Yundin V 2014 {\em Phys. Rev.\/} {\bf D89}
  034019 (\textit{Preprint} \eprint{1309.6585})

\bibitem{Cordero:2015hem}
Cordero F~F, Hofmann P and Ita H 2015  (\textit{Preprint} \eprint{1512.07591})

\bibitem{Badger:2013ava}
Badger S, Guffanti A and Yundin V 2014 {\em JHEP\/} {\bf 03} 122
  (\textit{Preprint} \eprint{1312.5927})

\bibitem{Bern:2013gka}
Bern Z, Dixon L~J, Febres~Cordero F, Höche S, Ita H, Kosower D~A, Maître D
  and Ozeren K~J 2013 {\em Phys. Rev.\/} {\bf D88} 014025 (\textit{Preprint}
  \eprint{1304.1253})

\bibitem{Gleisberg:2008fv}
Gleisberg T and Hoeche S 2008 {\em JHEP\/} {\bf 12} 039 (\textit{Preprint}
  \eprint{0808.3674})

\bibitem{Gleisberg:2007md}
Gleisberg T and Krauss F 2008 {\em Eur. Phys. J.\/} {\bf C53} 501--523
  (\textit{Preprint} \eprint{0709.2881})

\bibitem{Mastrolia:2011pr}
Mastrolia P and Ossola G 2011 {\em JHEP\/} {\bf 11} 014 (\textit{Preprint}
  \eprint{1107.6041})

\bibitem{Badger:2012dp}
Badger S, Frellesvig H and Zhang Y 2012 {\em JHEP\/} {\bf 04} 055
  (\textit{Preprint} \eprint{1202.2019})

\bibitem{Zhang:2012ce}
Zhang Y 2012 {\em JHEP\/} {\bf 09} 042 (\textit{Preprint} \eprint{1205.5707})

\bibitem{Kleiss:2012yv}
Kleiss R~H~P, Malamos I, Papadopoulos C~G and Verheyen R 2012 {\em JHEP\/} {\bf
  12} 038 (\textit{Preprint} \eprint{1206.4180})

\bibitem{Feng:2012bm}
Feng B and Huang R 2013 {\em JHEP\/} {\bf 02} 117 (\textit{Preprint}
  \eprint{1209.3747})

\bibitem{Mastrolia:2012an}
Mastrolia P, Mirabella E, Ossola G and Peraro T 2012 {\em Phys. Lett.\/} {\bf
  B718} 173--177 (\textit{Preprint} \eprint{1205.7087})

\bibitem{Mastrolia:2013kca}
Mastrolia P, Mirabella E, Ossola G and Peraro T 2013 {\em Phys. Lett.\/} {\bf
  B727} 532--535 (\textit{Preprint} \eprint{1307.5832})

\bibitem{Badger:2013gxa}
Badger S, Frellesvig H and Zhang Y 2013 {\em JHEP\/} {\bf 12} 045
  (\textit{Preprint} \eprint{1310.1051})

\bibitem{Badger:2015lda}
Badger S, Mogull G, Ochirov A and O'Connell D 2015 {\em JHEP\/} {\bf 10} 064
  (\textit{Preprint} \eprint{1507.08797})

\bibitem{Kosower:2011ty}
Kosower D~A and Larsen K~J 2012 {\em Phys. Rev.\/} {\bf D85} 045017
  (\textit{Preprint} \eprint{1108.1180})

\bibitem{Larsen:2012sx}
Larsen K~J 2012 {\em Phys. Rev.\/} {\bf D86} 085032 (\textit{Preprint}
  \eprint{1205.0297})

\bibitem{CaronHuot:2012ab}
Caron-Huot S and Larsen K~J 2012 {\em JHEP\/} {\bf 10} 026 (\textit{Preprint}
  \eprint{1205.0801})

\bibitem{Johansson:2012zv}
Johansson H, Kosower D~A and Larsen K~J 2013 {\em Phys. Rev.\/} {\bf D87}
  025030 (\textit{Preprint} \eprint{1208.1754})

\bibitem{Johansson:2013sda}
Johansson H, Kosower D~A and Larsen K~J 2014 {\em Phys. Rev.\/} {\bf D89}
  125010 (\textit{Preprint} \eprint{1308.4632})

\bibitem{Johansson:2015ava}
Johansson H, Kosower D~A, Larsen K~J and Søgaard M 2015 {\em Phys. Rev.\/}
  {\bf D92} 025015 (\textit{Preprint} \eprint{1503.06711})

\bibitem{Gluza:2010ws}
Gluza J, Kajda K and Kosower D~A 2011 {\em Phys. Rev.\/} {\bf D83} 045012
  (\textit{Preprint} \eprint{1009.0472})

\bibitem{Schabinger:2011dz}
Schabinger R~M 2012 {\em JHEP\/} {\bf 01} 077 (\textit{Preprint}
  \eprint{1111.4220})

\bibitem{Ita:2015tya}
Ita H 2015  (\textit{Preprint} \eprint{1510.05626})

\bibitem{Larsen:2015ped}
Larsen K~J and Zhang Y 2016 {\em Phys. Rev.\/} {\bf D93} 041701
  (\textit{Preprint} \eprint{1511.01071})

\bibitem{Cheung:2009dc}
Cheung C and O'Connell D 2009 {\em JHEP\/} {\bf 07} 075 (\textit{Preprint}
  \eprint{0902.0981})

\bibitem{Bern:2010qa}
Bern Z, Carrasco J~J, Dennen T, Huang Y~t and Ita H 2011 {\em Phys. Rev.\/}
  {\bf D83} 085022 (\textit{Preprint} \eprint{1010.0494})

\bibitem{Davies:2011vt}
Davies S 2011 {\em Phys. Rev.\/} {\bf D84} 094016 (\textit{Preprint}
  \eprint{1108.0398})

\bibitem{Bern:2008qj}
Bern Z, Carrasco J~J~M and Johansson H 2008 {\em Phys. Rev.\/} {\bf D78} 085011
  (\textit{Preprint} \eprint{0805.3993})

\bibitem{Bern:2012uf}
Bern Z, Carrasco J~J~M, Dixon L~J, Johansson H and Roiban R 2012 {\em Phys.
  Rev.\/} {\bf D85} 105014 (\textit{Preprint} \eprint{1201.5366})

\bibitem{Hodges:2009hk}
Hodges A 2013 {\em JHEP\/} {\bf 05} 135 (\textit{Preprint} \eprint{0905.1473})

\bibitem{Elvang:2013cua}
Elvang H and Huang Y~t 2013  (\textit{Preprint} \eprint{1308.1697})

\bibitem{Henn:2014yza}
Henn J~M and Plefka J~C 2014 {\em Lect. Notes Phys.\/} {\bf 883} 1--195

\bibitem{Dixon:1996wi}
Dixon L~J 1996 {\em {QCD and beyond. Proceedings, Theoretical Advanced Study
  Institute in Elementary Particle Physics, TASI-95, Boulder, USA, June 4-30,
  1995}\/} (\textit{Preprint} \eprint{hep-ph/9601359})
  \urlprefix\url{http://www-public.slac.stanford.edu/sciDoc/docMeta.aspx?slacPubNumber=SLAC-PUB-7106}

\bibitem{Mangano:1990by}
Mangano M~L and Parke S~J 1991 {\em Phys. Rept.\/} {\bf 200} 301--367
  (\textit{Preprint} \eprint{hep-th/0509223})

\end{thebibliography}

\end{document}